# NGC 6302: The Tempestuous Life of a Butterfly

Bruce Balick[1*], Lars Borchert[2], Joel H. Kastner[3], Adam Frank[4], Eric Blackman[5], Jason Nordhaus[6,7], & Paula Moraga Baez[8]

[1] Department of Astronomy, University of Washington, Seattle, WA 98195-1580, USA; balick@uw.edu
[2] Department of Physics and Astronomy, Aarhus University; Ny Munkegade 120, DK-8000 Aarhus C, Denmark; larseric@post.au.dk
[3] Center for Imaging Science and Laboratory for Multiwavelength Astrophysics, Rochester Institute of Technology, Rochester, NY 14623, USA; jhk@cis.rit.edu
[4] Department of Physics & Astronomy, University of Rochester, Rochester, NY 14627, USA; blackman@pas.rochester.edu
[5] Department of Physics & Astronomy, University of Rochester, Rochester, NY 14627, USA; blackman@pas.rochester.edu
[6] Center for Computational Relativity and Gravitation, Rochester Institute of Technology, Rochester, NY 14623, USA; jtnsma@rit.edu
[7] National Technical Institute for the Deaf, Rochester Institute of Technology, NY 14623, USA; jtnsma@rit.edu
[8] School of Physics and Astronomy and Laboratory for Multiwavelength Astrophysics, Rochester Institute of Technology, USA

## Abstract

NGC 6302 (The Butterfly Nebula") is an extremely energetic bipolar nebula whose central star is among the most massive, hottest, and presumably rapidly evolving of all central stars of planetary nebulae. Our proper-motion study of NGC 6302, based on excellent HST WFC3 images spanning 11 yr, has uncovered at least four different pairs of expanding internal lobes that were ejected at various times over the past two millennia at speeds ranging from 10 to 600 km s$^{-1}$. In addition, we find a pair of off-axis flows in constant motion at ~760 ± 100 km s$^{-1}$ within which bright [Fe II] "feathers" are conspicuous. Combining our results with those previously published, we find that the ensemble of flows has an ionized mass > 0.1 M$_\odot$. The kinetic energy of the ensemble, $10^{46} - 10^{48}$ ergs, lies at the upper end of gravity-powered processes such as stellar mergers or mass accretion and is too large to be explained by stellar radiation pressure or convective ejections. The structure and dynamics of the Butterfly Nebula suggests that its central engine has had a remarkable history, and the highly unusual patterns of growth within its wings challenge our current understanding of late stellar mass ejection.

*Unified Astronomy Thesaurus concepts:* Planetary nebulae (1249); Bipolar Nebulae (155); Ejecta (453); Stellar Wind Bubbles (1635); Stellar winds (1636); Shocks (2086)

## 1. Introduction

NGC 6302 (aka "The Butterfly") is an extraordinarily active and large bipolar planetary nebula ("PN") powered by an exceptionally energetic central source ("CS"[1]). The highly unusual morphological and dynamical complexity of NGC 6302 was first pointed out in the 1980s by Meaburn et al. (1980a,b "M+80a", "M+80b") who found compelling evidence of very fast stellar winds ~800 km s$^{-1}$ and internal motions of ~ ±180 km s$^{-1}$. Subsequent high-resolution images reveal that the geometric simplicity of the outlines of its lobes ("wings") obscure the highly complex structure and dynamics within their interiors As illustrated in Figure 1 and noted by Aller et al. 1981, NGC 6302 is a PN whose "appearance suggests violent motions with some bilateral symmetry".  Moreover, as we show later, the large mass of the ionized nebula, $M_{neb}$, stellar mass injection rate, $\dot{M}_\star$, stellar wind speed, $v_{wind}$, and the numbers of distinct flows of distinct ages in the wings are highly unusual among bipolar PNe. For these reasons, "sneezes and wheezes of a dragon" is an apt description of gas ejection history of the Butterfly.

This paper focusses on the patterns of proper motions changes in structure within the wings of NGC 6302—uncovered since Kastner et al. 2022 (K+22a,b) was published—and their consequences.  Those papers presented and discussed a suite of deep and panchromatic Hubble Space Telescope (HST) images of NGC 6032 taken with the Wide Field Camera 3 (WFC3) in 2019–20.  As K+22a noted, the nebula separates into a ~30″ inner core and two larger lobes, or butterfly "wings" to the E and W ("E" and "W" wings; Figure 1.) Other features that lie within the nebular core are largely peripheral to this paper.

---

[1] "CS" often denotes "central star" in papers like this one. We use CS to represent "central source" — a functional term that connotes a generic central engine that creates outflows of sundry mass flux, momentum, shape, and orientation—possibly intermittently.  We adopted "central source" in order to account for the complexity of the nebular structure and dynamics the central source may be a binary or multiple star system, possibly with internal mass flows and accretion disks.

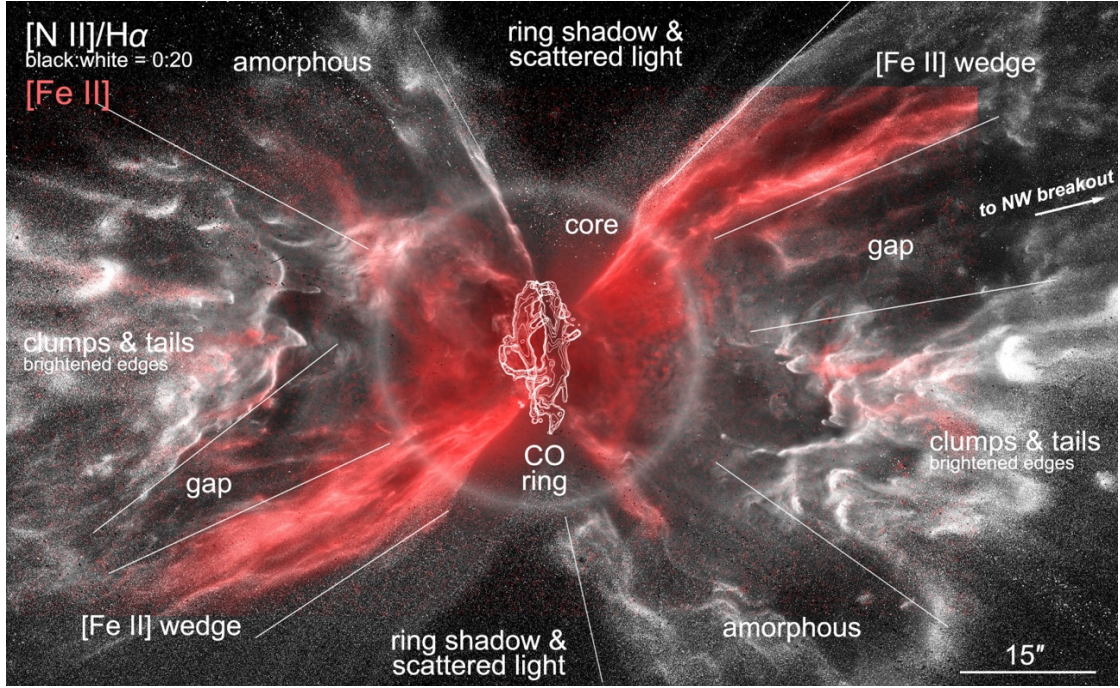

**Figure 1.** We show a slightly modified version of Figure 5 in K+22a that identifies the major wedges of NGC 6302 as seen in contrast-enhanced versions of [FeII] (red) and [NII]/H$\alpha$ (grey) images. Various wedge-shape zones beyond the core (grey circle) of radius 15″ are identified. The CO ring was copied from Santander et al. 2017, Figure 3. North is up.

Seen in projection, the morphologies of the structures found within the wings cleave into opposed pairs of wedges—an unprecedented property among bipolar PNe and first described in K+22a,b. For example, the pair of "[Fe II] wedges" contain radial "feathers" seen in the $\lambda 1.64$ μm shock-excited line of [Fe II]. The largest and optically brightest wedge pair consists of an array of edge-brightened clumps with downstream tails that give the nebula its "violent appearance" (cf. Aller et al. 1981). The "gap wedges" connect the HST images to the symmetry axes of the extended "northwest breakout lobe" ("NWBL") that extends beyond the WFC3 field of view (Figure 2a). A south-eastern counterpart to the NWBL may also be discernible.

The WFC3 images and the proper motion studies that comprise the foundation of this study are described is section 2, along with a short review of the key results of K+22a and previous papers. Section 3 explores the form, energetics, and the history of the winds that have shaped the wings. In the final section we trace some of the behaviors of the unusually massive CS of NGC6302 as it enters the final white dwarf ("WD") phase of its evolution. We review the challenges that the present study raises.

## 2. Data & Results

This section focusses on changes in structure within the wings of NGC 6302 between 2009 and 2020 as seen in the WFC3 camera and F658N images with 0.″0396 pixels. Calibrated images were extracted from the MAST Archive (the reader is referred to K+22a for a description of the images and their calibrations). The F658N filter passband is dominated by the bright nebular emission line of [NII] $\lambda 6583$Å, a line arising in ionization fronts ("I.F.s") or shocked gas low-ionization gas and that is found near the lobe edges of in most bipolar PNe. Our discussion also includes the [Fe II] image from K+22a (figures 2a and 2b) that is displayed using an intensity scale that emphasizes both its bright and faint features.  Hereafter we adopt a distance $D_{kpc}$ = 1 to NGC 6302 (K+22a, Gómez-Gordillo et al. 2020).

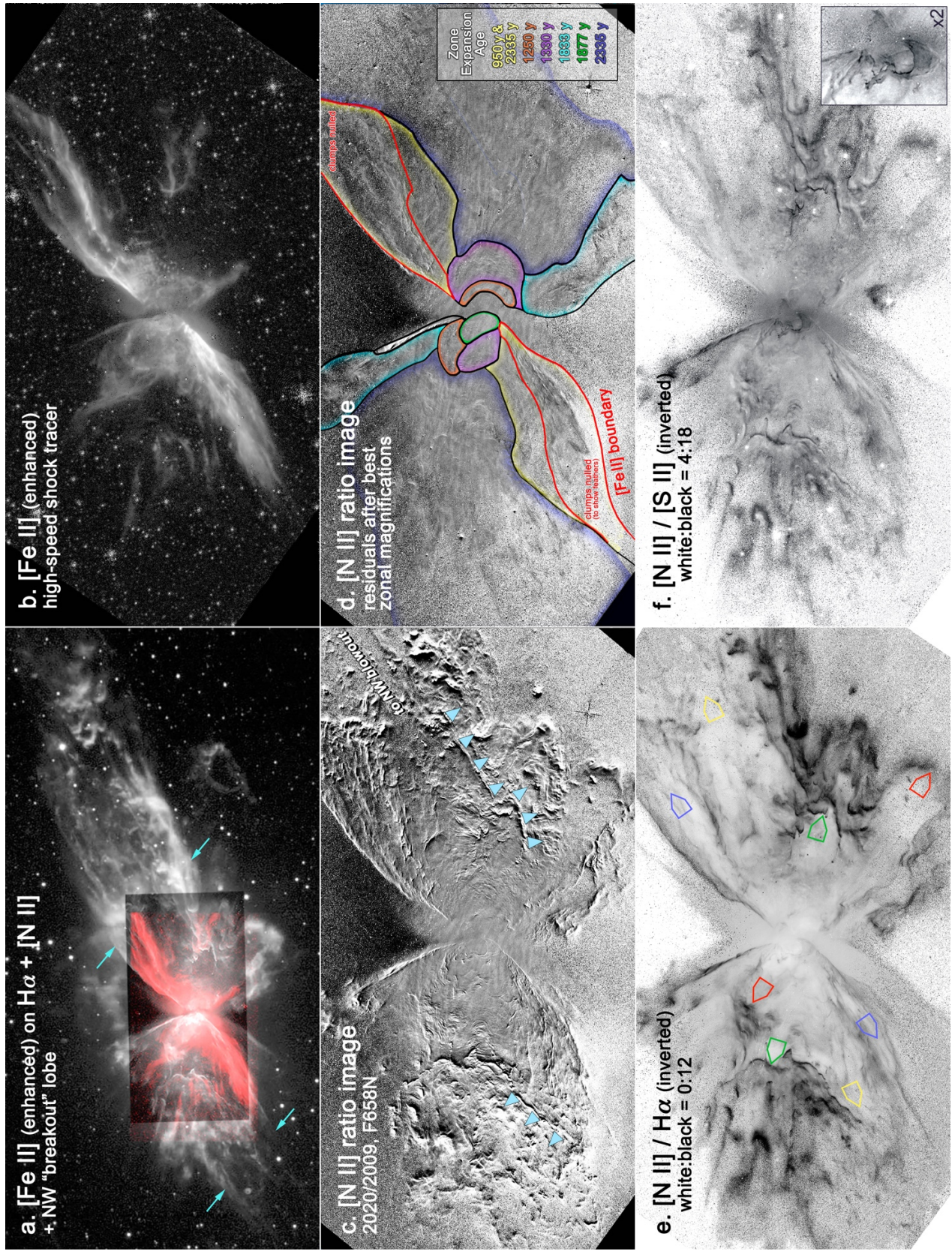

**Figure 2.** Sundry images of the structure (panels a & b), proper motions (panels c & d), and ratios of low-ionization images (panels e & f) of NGC6302. The grey-scale image in panel 2a is a deep wide-field Hα + [N II] image (esahubble.org/images/heic0407b/), taken by Romano Corradi (courtesy Albert Zijlstra). Figures 2b-f are aligned 1000x2000 portions of HST images and line ratios that are displayed at 0.″079/pixel. North is up.

*2.1 Proper Motions*

2.1.1. Methodology. Proper motion studies of the wings began with Meaburn et al. 2005; "M+0-5"). Our results were extracted from HST F658N images from 2009 and 2020 in which nebular features are sharpest and brightest. We initially applied the "difference method" (Reed et al. 1999) to trace their 11-year proper motions. "Residual images" were created from the differences and ratios of the original pair of images after the best magnification factor $M$ is applied to the 2009 images. This method determines radial proper motions from the two-epoch differences of the brightest features after finding and applying a magnification factor $M$ that, when applied to the earlier image, nulls the residuals. The two-epoch images were aligned by sub-pixel translations of the 2009 image relative to an expansion center ("EC") that was determined by trial and error. Referencing the proper motions to the EC balances the radial proper motions throughout the wings.

We also used a similar variant, the "ratio method", which the enhances the signal of radial motions of faint structures near the edges of the fields. If the nebula expands uniformly then, after magnification and subtraction or division, the residual image will be null other than noise, camera imperfections, and field stars.  Although the two methods yield equal values of $M$, we preferred the ratio method since signals of change are more readily apparent for outer structures of low surface brightness.  The residual of the ratio image is shown in Figure 2c prior to applying any magnification. Leading white and trailing black patterns on opposite sides of thin features are the usual signs of changes in location. White (black) pixels have values of 1.5 (0.5), so the signal of proper motion is very strong.  No changes in surface brightness were noted.

Ultimately, we found two superimposed and easily distinguishable types of zones of coherent motions in the wings of NGC 6302.  One type is the simple radial expansion with respect to the EC that the difference and ratio images presume ("ballistic motion") for deriving the optimum value of $M$.  The second type is a simple translation in which all features in the zone move in unison away from the EC (a steady radial flow).  We found only one such zone.  The [N II] features defining this flow lie within the [Fe II] feathers (figures 1 and 2b).

2.1.2 Zones of Coherent Radial Expansions ("ZRE").  We emphasize that a ZRE is a bounded zone, such as an ensemble of clumps, within which small features share a common magnification and expansion age.  The strings of cyan triangles indicate edges whose proper potions are primarily lateral rather than radial. The outermost edges of the wings also show slight lateral (circumferential) motions that suggest that the opening angles of the of the wings are increasing. See Appendix A for more details.

It quickly became clear that no single magnification factor nulls all of the structures within the wings. To explore the expansion patterns further, one of us (L.B.) made a movie of the residuals of ratio images in which $M$ increases in increments of 0.00143 from 1.0000 to 1.0160.  As shown in Table 1, we located several pairs of distinct ZREs, each of which is nicely nulled in the residual image by a zonal magnification, $M_{ZRE}$ (Table 1).  The approximate boundaries of the pairs of ZRE's are shown in Figure 2d using black lines and colored interior shading to identify them.  With one exception discussed in section 2.1.3, the ratio method very successfully nulled the residuals in each pair of ZREs (Figure 2d).

Values of $M_{ZRE}$ and their corresponding expansion ages $\Delta t_{ZRE} = 11\,\text{yr}/(M_{ZRE} - 1)$ are given in the first two columns of Table 1. The ages are independent of flow inclination. The Table also includes the axial speed at or near the outer extremities of the ZRE projected onto the plane of the sky, $v_{tip}$, assuming $D_{kpc}$

Table 1. Fitted parameters of the ZREs

| Best $M_{ZRE} - 1$ (uncertainty) | $\Delta t_{ZRE}$ (yr) | $\theta_{tip}$ (″) | $v_{tip}$ = tip extent/$\Delta t_{ZRE}$ (km s$^{-1}$ in sky plane) | Speed gradient ± 30% $v_{tip}$ (km s$^{-1}$)/$\theta_{tip}$ (″) |
|---|---|---|---|---|
| 0.00471 (10%) | 2335 | 177* | 360 | 2.0† |
| 0.00586 (20%) | 1877 | 45 | 113 | 2.7 |
| 0.00600 (10%) | 1833 | 5.2 | 13 | 2.9 |
| 0.00786 (20%) | 1330 | 20 | 68 | 3.5 |
| 0.00914 (10%) | 1250 | 15 | 60 | 4.0 |

* Tip of NWBL (Knot 1 of M+08)
† In good agreement with proper motion studies of knots by Meaburn et al. 2005 ("M+05") and equation (2) of Lago et al. (2019)

= 1, where $v_{tip}$ is derived from 4740 $(\Delta\theta/\Delta t)\, D_{kpc}$ (Reed et al. 1999), $\Delta\theta$ is its angular displacement in the time interval between images, $\Delta t$ = 11 yr. $\Delta\theta$ is given by $(M - 1)\,\theta_{tip}$, where $\theta_{tip}$ is the angular offset of the flow tip from the center of the core in arcsec. ($\theta_{tip}$ and $v_{tip}$ are especially uncertain since ZREs have ambiguous visual edges.) We cannot correct the values in Table 1 for the unknown inclinations of the flows associated with each ZRE.

Three pairs of the ZRE's are confined to the core region; i.e.; $\theta_{tip} < 20″$. The others align with radial wedge boundaries (Figure 1). As expected for bipolar PN, the ZRE's generally come in pairs with one exception—the green ZRE—whose western counterpart that is likely to be obscured by strong foreground dust extinction (K+22a). A large, subtle pattern of ripples caused by interference between the filter's reflective plates is also barely visible in all ZREs (Figure 2d).

The sharpest edges of features needed for proper-motion studies are also the brightest in the F658N image pairs where the [N II]/H$\alpha$ ([S II]/H$\alpha$) surface brightness ~6 (~0.4) (e.g., Rauber et al. 2014, "R+14"). Additionally, we checked the F658N results of proper motions in the zonal expansions against two-epoch F656N and F673N pairs, respectively. All outcomes are in good agreement, although the signal-to-noise ratio of the F658N images is superior to those of the noisier F656N and F673N images. For example, the uncertainties in the F658N residual images used here are half as large of F656N.

The proper motions derived from the [N II] image pairs agree, on the whole, with the independent estimates of the aggregate proper motions within the core made from the 2009 and 2020 WFC3 images of F673N by K+22a (Figure 11). The present results are also roughly similar to those earlier studies by Meaburn et al. 2008 ("M+08") and Szyszka et al. 2011, though with several key qualitative differences.

2.1.3. Zones of Coherent Translations ("Steady flows").

The residual of the ratio image in the [Fe II] wedge is not flat at any value of $M$; instead, it shows an ensemble of generally radial string-like residuals which shifts steadily as $M$ is increased (Figure 2d and Appendix A). We eventually found these sinews were nulled subtracting the image pairs after applying a small radial translation of the earlier image by $\Delta r$ = 4.0 ± 0.5 pixels. The translations correspond to constant radial speeds of 760 ± 100 km s$^{-1}$ in sky plane for $D_{kpc}$ = 1. If the leading edge of each flow lies at a projected angular radius of 45″ from the nucleus, then this flow began ~300 yr in the past.

The closest counterparts of the coherent translation feature in the [Fe II] wedges are some lie within the jets of Herbig-Haro Objects (Erkal et al. 2021), and the roughly radial strings of knots on the symmetry axis of the possible pre PN, Hen 2-90 (Sahai & Nyman 2000, Sahai et al 2002). However, the spatially resolved sinewy filaments that define the flow in the wings of NGC 6302 lie within an open wedge.

### 2.2 Key Features and Background Information

All of the major features in the wings of NGC 6302, such as the clumps, are visible in the each of the narrow filters other than [Fe II], albeit to different degrees (K+22a, Figure 2[2]). The star-facing sides of the clumps show locally enhanced [S II]/H$\alpha$, [N II]/H$\alpha$, and [N II]/[S II] line ratios (figures 2e & 2f) that are normally associated with dense I.F.s or the post-shock recombination zones (Figure 2f and Appendix B), where stellar winds or thermal expansion make contact with the surfaces of the clumps. We presume that that the star-facing sides of the brightest clumpy features are primarily ionized by stellar UV photons, though this is disputed by studies of Lago et al. 2019. On the other hand, sinewy filaments with nearly radial orientation, where stellar UV radiation impinges at very high obliquity and thus low flux, are excellent candidates for excitation by shocks. This is discussed later in the paper.

However, there is no doubt that [Fe II] is a standard tracer of relatively fast shocks (shock speed $v_s \gtrsim$ 50–100 km s$^{-1}$) since the line is frequently observed in the nebulae surrounding supernovae, dense high-speed bullets, and H-H Objects. The [Fe II] feathers argue strongly that stellar winds permeate the [Fe II] wedges of NGC 6302 and may have exerted a long-term hydrodynamical ("hydro") influence on their structure and evolution. This also applies to the sinewy [N II] features in the [Fe II] wedge whose proper motions are dominated by translations at speeds of about 760 km s$^{-1}$ (Section 2.1.3).

### 2.3 The Extraordinary NW Breakout Lobe

The Northwest Breakout Lobe ("NWBL") (Figure 2a) is one of the largest features of any PN. The NWBL extends well outside the wings of the Butterfly Nebula and beyond the field of view of the WFC3 camera. Using an ensemble of bright knots along the outer edges of the NWBL, M+05 and M+08 found that it lies close to the plane of the sky and extends 2′.95 from the nucleus (0.86 pc at $D_{kpc}$ = 1). They also found that the outer tip of the NWBL has a Doppler speed of ~600 km s$^{-1}$ and that the NWBL expands uniformly at ~0.5 mas yr$^{-1}$ per arcsec[3]. The lateral width of the NWBL expands at ~120 km s$^{-1}$ (M+05), as expected for aspect angle of this overall pattern of uniform growth.

M+05 obtained deep, high-spectral-resolution [N II] spectra of the NWBL through parallel E-W slits. Their slit 1 crosses the nebular nucleus, whereas slits 2, 3, and 4 cut through outer parts of the NWBL. All of the spectra in their slits have similar speed gradients, suggesting that the age of the entire NWBL is ~2200 yr (M+05, M+08) and that the dynamical origin of the NWBL lies at the center of the core.

A close look at our Figure 2a shows that the NWBL and the NW gap wedge are aligned and, so, may well be physically connected. This geometry is mirrored by a ridge of H$_2$ on one side of the gap wedge in the NW (Davies et al. 2003) and [Fe II] feather on opposite side of the base of the NWBL (Figure 2a). Figures 2c, e, and f, which are essentially maps of ionization and dynamical interfaces, also show this connection. (See also the SHAPE model in Figure 15 and the discussion in section 5 of M+05.)

---

[2] For later reference, note that the brightnesses of the lines of all ionization potentials reach their maxima at the same location just east of the nucleus. This is very difficult to reconcile with radiative ionization models (e.g., Osterbrock & Ferland 2006).
[3] This corresponds to about 2.5 km s$^{-1}$ per arcsec at $D_{kpc}$ = 1.0 after an inclination correction (M+05). Only Hen3–1475 has a higher Doppler growth rate among PNe: Borkowski & Harrington 2001, Fang et al. 2018a.

Red, yellow, green, and blue pairs of colored pointers in Figure 2e indicate other smaller pairs of seemingly hollow protuberances (or perhaps intrusions) which, like the breakout lobes, may have been produced by fast collimated ejecta from the nucleus. Aside from the flow into the [Fe II] wedge (blue pointers), all of these share the same expansion pattern and age as the NWBL. Thus, these protuberances in the clump wedges were also formed in the earliest ejection event.

The average expansion speed of the large and rapidly expanding NWBL, $\theta_{tip}/\Delta t_{ZRE}$, (M+05) as well as its mass (section 3.2), chemical composition (R+14), and shape (Figure 2a) seem similar to those other ZREs (Table 1). Therefore, given their similar ages, shapes, and proximity, we can presume that the NWBL and the clumps and gaps in Figure 1 are all part of a single eruption event, possibly the first in a series of ejection events (Figure 2d) that started ~2300 yr ago and became dormant ~1400 yr later. However, it is clear that the history of the "steady flow" in the [Fe II] zone that contains features moving in unison at ≈760 km s$^{-1}$ has had a different history.

*2.4 The Central Source*

Properties of the so-far undetected CS of NGC 6302 have been inferred from nebular observations and stellar models (see K+22a, section 6.2). Assuming that a single hot star dominates the nebular excitation, its temperature, 220,000K, and luminosity, 5700 – 11400 L$_\odot$, were derived by Wright et al. 2011 ("W+11") using a specific geometric density model and the ionization code "MOCASSIN". These properties place the CS at the top of the white-dwarf cooling sequence for stars of solar-metallicity in the H–R Diagram (Miller Bertolami 2016, "MB16", Figure 8). Based on their estimate of the present mass in the nebula and their adopted distance $D_{kpc}$=1.17, W+11 inferred that the initial mass of the CS, $M_{\star,init}$, is ~5.5 M$_\odot$. However, the location of the CS in the H–R Diagram and the adoption of a smaller distance, 1 kpc, led K+22a to suggest smaller values of $3.7 < M_{\star,init} < 4.8$ M$_\odot$. Newer models of the H-R tracks of post TP-AGB CS by MB16 imply even lower initial masses, at least for isolated stars.

**3. Discussion**

Nebular contents are shaped by the ram pressures of winds that reach their boundaries or by the thermal pressures that the hot gas that may have been thermalized in reverse shocks upstream. In this section we look more closely at the historical responses of the nebula to the history of the complex pressures of the outflows presumably powered by the massive and presumably rapidly evolving CS of NGC 6302. Since there is no evidence to support them, we shall ignore magnetic fields and pressures hereafter. Parts of this section are forensic and intrinsically somewhat speculative.

*3.1 Outflow Mass, Injection Rate, Speed, & Energetics*

Estimating the total mass of the ionized lobes, $M_{neb}$, is a key step to understanding the source of their energetics. W+11 estimated $M_{neb}$ ~1.82 M$_\odot$ based on a model with detailed radiation transfer, very specific geometry, a constant density of 2000 cm$^{-3}$, and $D_{kpc}$ = 1.17. This value of $M_{neb}$ scales to ~1.3 M$_\odot$ at $D_{kpc}$ = 1, as we adopt here. This estimate of $M_{neb}$ is plausible given the large initial mass of the CS of NGC 6302. However, their estimate of $M_{neb}$ depends sensitively on the density distribution, $n_e(r)$, since it scales as $1/n_e(r)$. R+14 (Figure 7) showed that $n_e(r)$ follows a complex radial density distribution from $10^{4.5}$ cm$^{-3}$ in the core to $10^2$ cm$^{-3}$ at the outer periphery 30″ E and W of the core.

A simple alternate method of estimating $M_{neb}$ derived by Pottasch (1984) is based on the nebular H$\beta$ luminosity and average density $n_e$. Using measurements of log(H$\beta$) = –10.55 (Tsamis et al. 2003), the same value of $n_e$, 2000 cm$^{-3}$ and $D_{kpc}$ adopted by W+11 we derive $M_{neb}$ = 0.59 M$_\odot$. If, however, we adopt our preferred value of $D_{kpc} = 1$, the average density from extinction-corrected [Ar IV]λ4711+ 4740Å lines from R+14, 8690 cm$^{-3}$, and He/H = 0.17 (R+14), then $M_{neb}$ becomes 0.12 M$_\odot$. This mass is consistent with recent values of $\dot{M}_\star \approx 10^{-4}$ M$_\odot$ yr$^{-1}$ and $\Delta t_{ZRE} \approx 2000$ yr. Of course, this mass is an underestimate since a fraction of the Hβ flux is heavily obscured by patchy dust, especially in the core (K+22a Figure 4). We shall assume that the ionized mass of the ZREs lies in the range 0.1 to 1 M$_\odot$ throughout this paper without trying to resolve the considerable disparities noted here. For comparison, Matsuura & Zijlstra (2005) find the mass in H$_2$ > 0.1 M$_\odot$, Trung et al. (2008) found a molecular mass in the core ≥ 0.5 M$_\odot$, and Matsuura et al. (2005) estimate a lobe mass of 0.4 M$_\odot$ from dust extinction.

Next, we review estimates of the flow speeds and mass injection rates from the CS. From their early imaging and spectroscopic observations of the NWBL M+80a,b found $v_{wind}$ ~ 600 km s$^{-1}$ at the lobe tips, 800 km s$^{-1}$ in the stellar winds; all in good accord with our proper motion results. They also concluded that the collimated stellar winds have excavated a pair of bipolar cavities "with flowing walls delineated by shocks as described by Cantó (1979)"—in good agreement with our findings in section 3.3 below. Finally, M+80a,b, Matsuura et al. (2005), M+08, Trung et al. 2008, Szyszka et al. 2011, Santander-García et al. 2017, and Bollen et al. 2020) all estimated $\dot{M}_\star \gtrsim 10^{-4}$ M$_\odot$ yr$^{-1}$ (at $D_{kpc}$ ~ 1). As noted by K+22a, "this [value of $\dot{M}_\star$] would far surpass the mass loss rates typically observed for even the most luminous AGB stars (Höfner & Olofsson 2018)".

In addition to mass injection, the CS is the overwhelming source of kinetic energy and heating in the wings. Nebular heating is generally dominated by stellar UV photons (Dopita & Sutherland 1996), and we can assume that the same is true throughout NGC 6302 except locally in shocks.

The total kinetic energy within the nebular wings is $E_{neb} = 1/2\, M_{neb}\, v_{neb}^2$, where $v_{neb}$ is the speed of the complex network of internal flows. If we adopt conservative representative values for $M_{neb} = 0.1$ M$_\odot$ and $v_{neb} = 100$ km s$^{-1}$ then $E_{neb} = 10^{46}$ erg. $E_{neb}$ increases to $10^{48}$ erg if we adopt $M_{neb} = 1.3$ M$_\odot$ (W+11) and $v_{neb} = 268$ km s$^{-1}$ as measured at point A′ near the midpoint of the NWBL (M+05). The total flow energy released by the CS can be even greater if winds escape through gaps in the wings. In any case, the kinetic energy carried within the nebular wings NGC 6302 is definitely impressive.

### *3.2 The art of reconstructing history*

We now begin a lengthy discussion aimed at understanding the dynamical history of the wings and the CS. First, note that reverse shocks can be locally important in converting the kinetic energy of fast stellar winds into thermal pressure. Hot gas quickly fills the volume between the reverse and leading shocks. In the case of round and elliptical PNe, this trapped hot gas forms a thermally and supersonically expanding "hot bubble" ($T_{bub} > 10^7$K) between the inner (reverse) shock and the leading (outer) shock just inside the displaced older gas (e.g., Kwok et al. 1978, Toalá & Arthur 2014 "TA14"). The soft X-rays sometimes detected near the edges of their soft-X-ray bubbles confirm their hot temperatures and large thermal pressures (Kastner et al. 2012, Toala & Arthur 2018).

We first consider some simple examples of the roles of ram and thermal pressures in shaping bipolar PNe. Early hydro simulations of bipolar PNe by Icke et al, (1989, 1992) showed that the lobes of bipolars can be formed solely by thermal pressure, much the same paradigm as that for round and elliptical PNe but with the expansion of the hot gas confined by a dense equatorial torus in the core. On the other hand, the lobes of young bipolar PNe are shaped, at least initially, by the ram pressure of fast collimated stellar winds, as first emphasized by Lee & Sahai (2003). They showed how collimated jets (i.e., ram pressures) injected into cold AGB winds will also produce lobes of appropriate shapes in young, neutral PNe. Huarte-Espinosa et al. (2012) showed how thermal pressure from subsequent UV heating could modify such lobes into the familiar shapes that we see in mature bipolar PNe.

However, NGC 6302 is not simple. The recurrent and misaligned flows within its wings tell a far more complex story. Unlike lobes of most bipolar PNe, images of the wings leave an impression of the violent history noted by Aller et al. We shall show that the interiors of the wings have been shaped and reshaped by the mass flows in various orientations. Also, the inertial resistance of the ensemble of interior clumps has played an important dynamical role in the evolution if gas within these wings.

The point is this: any discussion of the complex history the wings of NGC 6302 must take the roles of internal resistance, both ram and thermal pressures, and their sources, into consideration when talking about their histories. A large dose of forensic guesswork is needed in order to connect the dots and to form a coherent picture. Assembling such a scenario will take us through the remainder of the paper.

### 3.3 The changing roles of ram and thermal pressure

The momenta of collimated flows probably gave the basic form that we see now in the wings. Thermal pressures rise very rapidly once UV heating starts (Huarte-Espinosa et al.). At that point, reverse shocks retreat inwards towards the wind source (Balick et al. 2019). As a consequence, the hot interiors of the intrusions start to expand laterally, shocking and compressing the denser gas that surrounds them, as they do in along the bubble edges of normal round and bipolar PNe. So, thermal pressures are likely to influence and eventually dominate the gas dynamics within lobes and protrusions, as seems to be the case today for the NWBL and ZREs. Of course, this process repeats with each stellar ejection.

Is sharp contrast, M+05 favored shaping by a steady wind (e.g., Barral & Cantó, 1981). However, the survival of a steady off-axis stellar wind that persists for >1000 yr in a nebula with a very complex ejection history seems improbable. Moreover, any density inhomogeneities encountered by wind streamlines will disorient the flow and lead to the rapid formation of turbulence and internal shocks where wind thermalization rapidly occurs. For this reason, we expect that thermal pressure is extremely important in the wings except for the [Fe II] feathers that are being penetrated by a steady fast low.

M+05 alternatively suggested that the NWBL formed in a single ballistic ejection ~2300 yr ago. This type of ejection doesn't explain the non-radial motions along its lateral edges or the persistence of the dense and rapidly recombining [Fe II] feathers. A hybrid possibility may apply to the ballistic knots along the edges on the NWBL if these knots were initially formed by flow instabilities and launched ballistically early in the history of stellar mass ejection. This ejection pattern is portrayed in hydro simulations as a light, diverging fast stellar wind that strongly interacts with a denser AGB wind and forms surface instabilities that are quickly compressed into knots. The knots have higher specific momentum than their surroundings, so they diverge ballistically right after they are formed.

Consider the network of intrusions marked with colored pointers in Figure 2e. The intrusions may have been initially burrowed by collimated winds, but they may have become rounded as thermal pressure fills their interiors. Note that in NGC 6302, their shapes may also have been distorted through lobe-lobe interactions, thus somewhat obscuring their original forms. In contrast, the interiors of the [Fe II] feathers are presently dominated by the ram pressure of a steady and highly supersonic flow (section 2.1.3). Hydro models of the lobes formed by steady fast winds nicely match those of several simple bipolar PNe (e.g., Barral & Cantó, Lee & Sahai, Balick et al. 2018, 2019). However, the morphologies and kinematics of the feathers of NGC 6302 are poorly fitted by these models, all of which assume of winds speeds ≲400 km s$^{-1}$ and predict hollow lobes with a speed pattern that rises with radius. Instead, the feathers resemble the fast, steady-speed outflows of R Aqr (Liimets et al. 2018, figures 3 & 8).

Knots with tails are common in the wings of NGC 6302. The tails provide additional evidence for the ongoing importance of thermal pressures throughout the clump wedge. That is, expanding hot gas (i.e., thermal pressure) in which the knots are immersed will form and subsequently sustain outward-facing ablation tails (Steffen & López 2004, Toalá & Arthur 2011, "TA11", TA14).

### *3.4 Locating pressure interfaces.*

It is well known that winds and thermal pressures can form compressed and radiative (i.e., rapidly cooling) shock features at their interfaces. The types of shocks that form at these boundaries have shapes and emission-line spectra that reveal their nebular geometries, motions, and shock speeds. Thus, interfaces are also excellent tools to map the current large-scale dynamics within the wings.

However, both I.F.s and shocks can have similar appearances. Both are the loci where local pressures compress gas to high densities. The interfaces in the wings show up clearly as thin enhancements of [N II]/H$\alpha$ and [N II]/[S II] (figures 2e, & f; see also Appendix B for an explanation of the [N II]/[S II] ratio as a high-density tracer). The latter ratio is especially useful where in high-ionization regions where the [N II]/H$\alpha$ ratio may be only marginally detectable; i.e., $N^+/N \lesssim 0.01$. The proper motions of the interfaces can distinguish shocks from I.F.s: I.F.s quickly slow to the local sound speed, ~10 km s$^{-1}$ (e.g., Dyson & Williams 1980, Draine 2011) whereas active shocks maintain their initial speeds as wind momentum is deposited or as thermal pressures are sustained. Edges with large (supersonic) proper motions (Figure 2c) are excellent candidates for shocks since proper motions >20 km s$^{-1}$ are readily uncovered from our two-epoch studies.

Shock have higher excitation temperatures than I.F.s, so the two can often be distinguished by their ratios of temperature-sensitive emission lines. Unfortunately, there are no published spectroscopic observations along the bright that can be used to this end. R+14 mapped line intensities through some of the edges of the clumps; however, they don't report the actual values of key line ratios along these edges. Using the line intensity maps of R+14, L+19 asserted the significance of shock heating in some of regions of low ionization in the E-W direction from the center, but they didn't identify any specific shock loci nor did they attempt to estimate shock speeds, $v_{shock}$. Moreover, in reaching their conclusions they adopted the lowest value for the uncertain value of $L_\star$ from W+11, 5690 L$_\odot$. Their conclusions about shock-heating are not verifiable.

### *3.5 [Fe II] and H$_2$ Structure and Excitation*

[Fe II] is cited as an excellent tracer of wind-driven shocks in H-H Objects (Erkal et al.), dense bullets (e.g., Bally et al. 2020), and many supernova remnants in which the wind speeds are typically ≥100 km s$^{-1}$ (see Section 5.2 of K+22a). The pair of [Fe II] feathers of NGC 6302 (Figure 2b) are especially notable because the presence of [Fe II] lines in PNe is unusual[4] and because the steady fast flow of 760 km s$^{-1}$ that currently runs through them (section 2.1.3).

Oddly, the feathers lie in opposition along the outer NW and SE edges of the wings. Thus, the agent that excites [Fe II] follows an asymmetric course relative to the symmetry axes of the wings and the equatorial torus. This renders the "S"-shaped symmetry of the most prominent [Fe II] feathers (Figure 2b) as well as the constant, high-speed speed of the flow through the interiors of the [Fe II] wedges, distinct from the other wedges.

---

[4] This may simply be the result of only reporting detections.

Perhaps curiously, these [N II] filaments do not have counterpart in the [Fe II] images. If the discussion by Contini et al (2009) is applicable, then the local electron density $n_e \geq 10^6$ cm$^{-3}$ in the [Fe II] emission zone. This density is well above the critical density, $n_{crit}$, for the [N II] lines, where $n_{crit}$ is the density at which collisionally excited lines such as [N II] and [S II] are de-excited at equal rates by radiative emission and collisional thermalization (Appendix B). Therefore, the filaments are seen exclusively in [N II] or [Fe II] (figures 2a & 2e), but not both. Moreover, if $n_e \geq 10^6$ cm$^{-3}$, then the corresponding Fe$^+$ recombination time $\approx (10^5$ yr$)/n_e \approx 1$ month. Thus, the steady fast collimated winds of ~760 km s$^{-1}$ are presently actively shocking and shaping the gas the [Fe II]-emitting gas within this wedge.

Bright and nearly radial "ridges" of shocked H$_2$ $\lambda$2.122µm emission lie along the interfaces between the gaps and clumps in both the E and W wings of in NGC 6302, as noted in section 2.3. No [Fe II] is observed in the H$_2$ ridges, and no H$_2$ is seen anywhere that [Fe II] arises. This is no surprise: H$_2$ and [Fe II] lines are excited at complementary shock speeds and zones of ionization. We surmise that the H$_2$ is shock excited in NGC 6302 since most of the stellar UV reaching the ridges is diluted by the large obliquity of the ridge relative to radial lines from the CS. Yet, it seems improbable that winds from the CS reach these ridges directly. Instead, the shocks may be induced by hot gas flowing along the ridges.

The H$_2$ ridges in the wings share the same geometry and oblique stellar UV illumination as the H$_2$ that lines the lateral boundaries of the lobes of several bipolar PNe (Kastner & Weintraub 1996, Fang et al. 2018b). All of them are probably excited in the same manner.

## 4. Conclusions

In the previous section we made a forensic case that the CS shaped the wings of NGC 6302 in a series of metaphorical sneezes and a protracted wheeze, akin to a fire-breathing dragon. Next, we summarize our major findings and explore paradigms for the nature of the engine.

*4.1 Synopsis*

The wings of NGC 6302 are filled by pairs of outflow structures, each with its own expansion age and orientation. Generally speaking, the edges of these features that face the nucleus are very prominent in low-ionization emission lines, especially [N II] which is locally >ten times brighter than H$\alpha$. These edges are also outlined by strongly locally enhanced [N II]/[S II] by factors $\approx$ 4–5, which we interpret as the result of local quenching the [S II] lines where densities $\gtrsim 10^4$ cm$^{-3}$. We surmise that these edges are a combination of I.F.s photoionized by UV radiation from the CS and shocks driven either by winds or pressure and supersonic expansion of hot gas heated by thermalized winds from the CS. In, bright [Fe II] lines arise in a pair of conspicuous "feathers" that are seen adjacent to the lateral edges of the wings. The [Fe II] line is likely to be produced in active shocks at speeds >100 km s$^{-1}$ by a steady high-speed outflow containing [N II] filaments with proper motions of 760 km s$^{-1}$.

This scenario of the formation and shaping history of the wings emerges:
1. *~2300 yr ago.* The first and most energetic of a series of fast collimated outbursts from the CS formed the NWBL, several small radial intrusions that are marked in Figure 2e, and pairs of large and irregular groupings of clumps about 30″ E and W of the nucleus. All of these features show identical radial patterns of proper motions of thin features found along their edges—i.e., they grow uniformly in length and width and share the same kinematic ages.
2. *Between ~1200 and 1900 yr ago.* At least three weaker outburst events from the CS produced smaller coherent radial ejections but along different symmetry axes. These outer reaches of these outflows presently lie within the orange, green, and purple boundaries of Figure 2d. The "green" zone in the high-ionization heart of the core has an especially complex structure (inset Figure 2f).

3. *~300 yr ago to the present.* A fast and ongoing ejection event within the [Fe II] wedges is forming a bright pair of "[Fe II] feathers" within which a network of sinewy, almost radial features seen in [N II] move radially in unison at 760 km s$^{-1}$.

These events are seemingly coincided with the evolution of a putatively massive, luminous and very hot central star as it traversed the H–R Diagram. However, as we discuss next, the evolution of such a star may not be the entire story since companion stars or accretion flows probably played key additional roles in the mass ejection history into the wings.

*4.2 Formation and Ejection Paradigms*

It is clear that one-time ejection events (such as binary-star mergers and common-envelope ejections; García-Segura et al. 2022 and earlier papers in the series; Glanz & Perets 2021a, Zou et al. 2020) or episodic ejections with the period of a stable interacting binary aren't fully consistent with the proper motions within the nebular wings of NGC 6302.

By analogy to the formation of the homunculi of $\eta$ Carinae (Davidson, 2020) and X-ray bursters (e.g., Güver et al. 2022), the most likely source of the kinetic energy of the outflows, $\geq 10^{46}$ ergs, is gravitational infall[5]. Soker & Kashi 2012 ("SK12") and K+22a discussed the possibility of an "Intermediate-Luminosity Optical Transient" event in NGC 6302 in which mass from an AGB donor falls onto a companion main-sequence companion. SK12 describe the attributes of lobes launched by ILOTs. To summarize, the lobes must come in opposite pairs, each lobe must show a linear speed-radius pattern (i.e., linear P-V Diagram), the ejection occurs on a time scale of a few months at most, the speed of the outflow the escape velocity of the main-sequence star that ejects them (or slightly less if the flow has been decelerated by slower AGB gas downstream), and a total kinetic energy of $10^{46}$ to $10^{49}$ ergs. The only lobe pair that meets these requirements and has an appropriate kinematic age is the NWBL and its possible but marginally observable SE counterpart. The tip speed of the NWBL, ≈600 km s$^{-1}$, matches the escape speed of the Sun. Thus the companion star has a mass ~1M$_\odot$. However, an ILOT event would have ended long before the more recent flows had formed.

No matter how they were ejected, a third nearby companion star is needed to stimulate sporadic flow surges and to wobble the outflow source (e.g., an accretion disk). The importance of the shaping by tertiary stars is just starting to be appreciated (Bear & Soker 2017, Glanz & Perets 2021b, Hamer et al. 2022, see also De Marco et al. 2022).

Other types of paradigms are not very successful at accounting for the full range of observed behaviors in the wings of NGC 6302. For example, Betelgeuse may be an analogue since its surface is highly convective, having produced recurrent surface mass ejections ("SME"; Dupree et al. 2022). However, the ejection speed of an SME will be at or just above the surface escape speed, ≈15–20 km s$^{-1}$ (Jadlovský et al. 2023). Therefore, this mechanism is much too feeble to explain the much faster and more energetic outflows of NGC 6302.

It is also conceivable that the ejections of the ZREs of NGC 6302 are the result of recurrent and deep thermonuclear flashes. Certainly, the huge He/H and N/O ratios in the nebula (W+11, R+14, and papers cited therein) attest to exceptionally energetic nuclear activity such as hot bottom burning in at least one of the stars in the CS in the past (Ventura et al. 2018; see also the comprehensive review in Kwitter & Henry 2022). Even so, it is far from clear how the mass and energy of material released from core flashes leads to the formation and ejection of flows with the requisite collimation, speed, mass, mass loss rate, orientation, and cadence of a few hundred yr.

---

[5] For reference, the gravitational potential energy, GM/ΔR, of a 5-M$_\odot$ binary pair separated by ΔR = 1 AU is $10^{47}$ erg.

These or any other sort of explanations for the mass ejection of NGC 6302 will likely remain in the realm of speculation without better observational constraints on the so-far invisible CS. About all that we can state with reasonable confidence is that the star in the central engine that currently provides the bulk of the nebular heating has a luminosity lies between $10^{2.9}$ and $10^{4.4}$ L$_\odot$ (K+22a). Our future observational challenge is to determine the location of the most luminous star in the core of NGC 6302 on the H-R Diagram as well as those of its close companions.

### *4.3 Final Remarks*

Our findings set the CS of NGC 6302 and its nebula apart from other PNe that have been as carefully studied. Projecting forward another millennium, NGC 6302 may well morph into a sibling of NGC 2440, which has about the same ionization structure, multiple lobes of different ages and orientations (Lago & Costa 2016, "LC16"), N/H enrichment, radial density structure, and a central engine of similar initial mass, temperature, and luminosity, and physical size (0.6 pc) at $D_{kpc}$=1.8 but is expanding 50% more slowly (Hyung & Aller 1998, López et al. 1998, Bernard Salas et al. 2002, LC16, K+22a). The narrower NE-SW lobe pair has a tip speed ≈ 170 km s$^{-1}$ and expansion age of 1700 yr. We identify them as the evolutionary descendants of the NWBL. The E-W bulbous pair expands at about half that speed and an age of 3300 yr (see the SHAPE model and Table 1 of Lago & Costa).

Two other unusual non-bipolar PNe, NGC 3132 and 6210, must be mentioned. A very detailed study spectroscopic and proper motion study of NGC 6210 by Henney et al. (2021) revealed at least five distinct ejection axes with different orientations and ages over the past ~3500 yr. They propose that the nebula was shaped by mass transfer within a triple star system. For different reasons, much the same conclusion was reached for NGC3132 by De Marco et al.

Acknowledgements: The proper motion results in this paper are based on observations made with the NASA/ESA Hubble Space Telescope, obtained at the Space Telescope Science Institute (STScI), which is operated by the Association of Universities for Research in Astronomy, Inc., under NASA contract NAS5-26555. The 2019-20 (2009) observations are associated with program GO15953 (GO11504). JHK and PMB acknowledge support provided by the U.S. National Science Foundation ("NSF") grant AST-2206033 and NASA/STScI grant HST-GO-15953.001-A to RIT. AF acknowledges the support US Department of Energy grants DE-SC0001063 and DE-SC0020432. EB acknowledges support NSF grants AST-1813298 and PHY-2020249. JN acknowledges support from grant NSF AST-2009713.

This paper celebrates the consistently prescient imaging and high-dispersion spectroscopic studies of Prof. John Meaburn throughout his career. The observations of NGC 6302 by Prof. Meaburn and his collaborators (M+80a, M+80b, M+05, M+08) serve as a poignant and typical example of his excellence in the acquisition, analysis, and insightful interpretation on this object, other PNe, and similar nebulae.

**ORCID IDs**

Bruce Balick https://orcid.org/0000-0002-3139-3201
Joel H. Kastner https:/orcid.org/0000-0002-3138-8250
Adam Frank https:/orcid.org/0000-0002-4948-7820
Eric Blackman https:/orcid.org/0000-0002-9405-8435
Jason Nordhaus https:/orcid.org/0000-0002-5608-4683
Paula Moraga Baez https:/orcid.org/0000-0002-1042-235X

## Appendix A. Proper motions as tracers of expansions and steady flows

Panels a, b, and c in Figure A.1 shows ratios of the 2020 and 2009 F658N images on the eastern side of NGC 6302 for the three indicated values of the magnification factor *M*. The ratios range from 1.5 (white) to 0.5 (black) in all panels. Panel d is a difference image made after the 2009 image is displaced by 4.0 pixels along the symmetry axis of the zone outlined in red. (The results for the W wing are very similar.) The red lines outline the portions of the SE "[Fe II] feather" where the surface brightness of [Fe II] reaches its peak (Figure 2b). Note the sinewy radial filaments within this "red zone" in all of the panels.

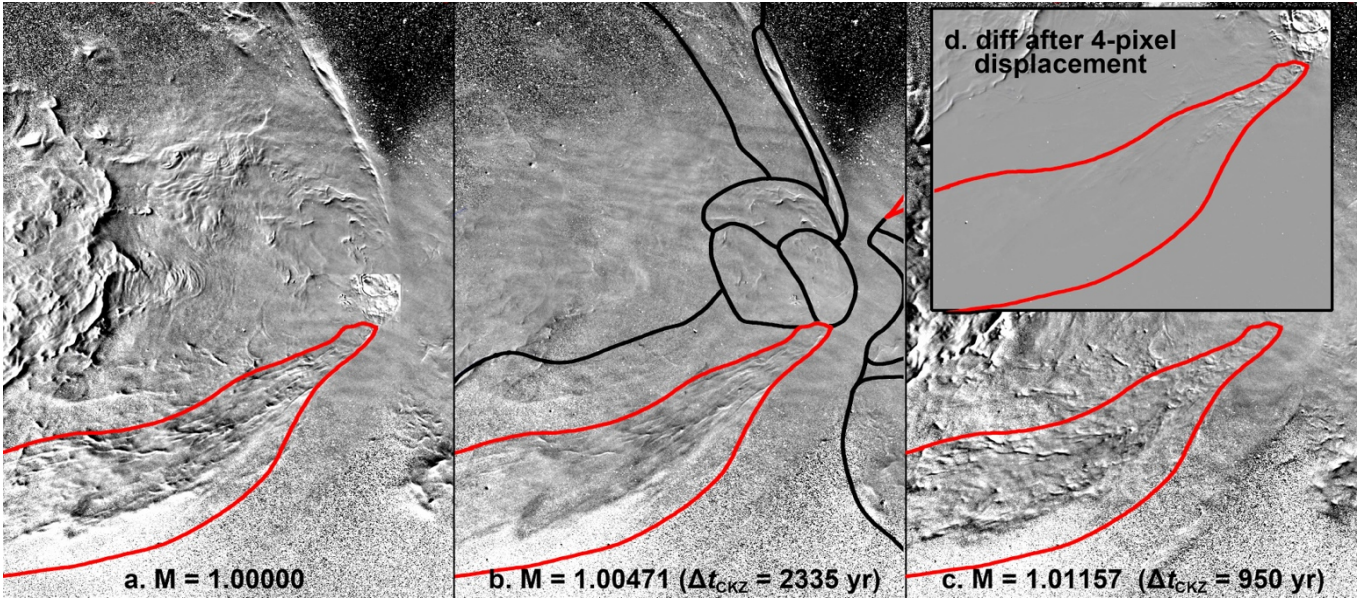

Figure A.1, panels a,b,c: The residuals of the 2020 and 2009 F658N image ratios after the latter has been magnified by a factor of *M*. Those features that share a common magnification (aka expansion age) are nulled if *M* is successfully chosen; however, no value of M nulls the full set of sinewy feauters in the red zone. Panel d: Residuals after a 4.0-pixel radial displacement of the 2009 image (see text).

Panel a illustrates the impressive strength of the raw signal of the proper motions. Panel b shows the residuals in the clump zone after magnifying the 2009 image by $M = 1.00471$ and dividing. The edges of the very bright clumps in their wedge zone are completely nulled. (Faint grey ripples across the residual image are instrumental in origin.) The boundaries between various ZREs are shown as solid black lines (as in Figure 2d.). In panel c we apply a large value of *M* in order to highlight features that were not nulled in the red zone.

There is no value of *M* for which the entire set of sinewy [N II] features is nulled in the red zone. However, the set of sinews is very successfully nulled by translating the 2009 image in the radial direction by $4.0 \pm 0.5$ pixels and subtracting (panel d). The same result applies to the [N II] sinews in the W wing as well. Thus, the sinewy features recede from the central engine at the same projected speed, $760 \pm 100$ km s$^{-1}$, at a distance of 1 kpc.

**Appendix B. The F658N/F673N ratio as a tracer of high densities**

As noted in Section 3.4, thin, dense rims of gas form where winds encounter, compress, and shock slower gas downstream. The loci of such rims reveal these interfaces, and their proper motions reveal the direction of the motions resulting from the pressure differences across the shock. The [N II]/H$\alpha$ (Figure 2e) ratio is one way to locate these thin rims. Here we discuss a novel yet highly effective display based on the F658N/F673N ratio images (Figure 2f), where the emission in the two filters is dominated by [N II]$\lambda$6584Å line and the sum of the two [S II] $\lambda$6717Å and $\lambda$6731Å lines, respectively.

We constructed the [N II]/[S II] ratio image with the prior expectation that the N$^+$/S$^+$ lines would arise in nearly identical volumes with the same abundances, ionic excitation temperatures, internal extinctions, and volume filling factors. In this case, the ratio image will show only very small-scale amplitude variations from its average value, and then only in thin loci along the outer edges of I.F.s or thin shocks[6]. This expectation is supported by excellent ground-based ionic maps in the two lines published in R+14, figures 2 & 4. However, this expectation isn't valid for the HST ratio images with superior spatial resolution shown in NGC 6302 (Figure 2f).

Curiously, the brightest N$^+$ and S$^+$ lines are found where *all* of the narrowband HST images, including He$^{++}$ (Noll, HST GO 11504) and [Ne V] (K+22a) attain their respective peak surface brightnesses. Ionization models predict that the fractional ionization of N$^+$ and S$^+$ should be negligible in the innermost core regions of NGC 6302 in zones of bright He$^{++}$ and [Ne V] where an extremely hard stellar UV radiation spectrum dominates, in accord with the estimated effective temperature of the ionization source, $T_{\text{eff}} \geq 200{,}000$ K. In such zones, N$^+$/N and S$^+$/S will be $\ll 10^{-2}$ (Osterbrock & Ferland 2006, Figure 2.6), so lines of N$^+$ and S$^+$ should be marginally observable in the core at best. This argues that both UV ionization and shock-compressed gas contribute to the emission in this zone. (Recall that Barral et al. 1982 and R+14 found that the observed density near the core is $10^{4.5}$ cm$^{-3}$ so that the recombination times of N$^{++}$ and S$^{++}$ $\leq$ 1 yr).

We attribute the small-scale and prominent enhancements of F658N/F673N in Figure 2e to a combination of these locally large densities in the thin post-shock recombination zones as well as the large spread in critical densities, $n_{\text{crit}}$, of the lines in the [N II]$\lambda$6584 and [S II]$\lambda$6717 and $\lambda$6731 lines. For the [S II] 6717Å and 6731Å lines of S$^+$ (the two $^2$D to $^4$S transitions), $n_{\text{crit}}$ is 1500 and 16,000 cm$^{-3}$, respectively, whereas $n_{\text{crit}}$ is 77,000 cm$^{-3}$ for the 6548Å and 6584Å lines ($^1$D$_2$ to $^3$P transitions)[7]. The spread of the values of $n_e$ and $n_{\text{crit}}$ imply that the F658N/F673N ratio image will vary considerably between $\approx 10^{3.5} < n_e < \approx 10^{5.5}$ cm$^{-3}$.

---

[6] For I.F.s, these loci are where S+ can be formed by sub-Lyman UV photons that penetrate into a fully recombined N zone. In the case of shocks, Raga et al. 2008 (Figure 3) show that the [S II] emissivity extends beyond that of [N II] along the recombination zone at the leading edges of bullets with pre-ionized shocks, especially if UV radiation from a hot central star is blocked by the bullet. See Hartigan et al. 1994, Riera et al. 2006, and Raga et al. 2008 for simulations of emission lines in shocked cloudlets (aka clumps).

[7] Values of $n_{\text{crit}}$ are taken from Tables 18.1 and 18.2 of Draine 2011.